\documentclass[12pt]{article}
\usepackage{graphicx}

\begin{document}

\title{Central configurations of the 4-Body Problem \\ with gravitational charges of both signs}

\author{E. Pi\~na and P. Lonngi\\
Department of Physics \\ Universidad Aut\'onoma Metropolitana - Iztapalapa, \\
P. O. Box 55 534, Mexico, D. F., 09340 Mexico \\
e-mail: pge@xanum.uam.mx \quad plov@xanum.uam.mx}

\date{ }
\maketitle

\abstract{
The Newtonian Four-Body Problem with positive and negative gravitational charges is studied from a physical point of view that considers inertial masses that are always positive and positive or negative charges in the gravitational interaction forces. The difference with respect to the differential equation of motion used in several mathematical papers is stressed. Formulation of the modified Dziobek equations is developed. We prove that four masses with one charge of different sign lead to a concave central configuration with the charge of different sign in the convex hull of the other three. For two charges of one sign and two charges of the opposite sign, we prove that the central configuration is convex with the two charges of the same sign on each diagonal. An algorithm used for computing the six distances between particles of a planar Four-Body central configuration, presented previously for positive gravitational masses, is adapted to compute physical central configurations with charges of different sign.
}
\


\


\newpage

\section{Introduction}
In this paper we present the study of central configurations of four particles that obey Newton's three laws of motion and Newton's gravitational force law \cite{tm}, \cite{js}, assuming the possibility of gravitational charges of both signs, according to the prescription (which is opposite to that in electrostatics for electric charges where G would have a negative value) that the force between two masses with charge of the same sign is attractive, while the force between two masses with charge of opposite sign is repulsive. More explicitly, we assume the equations of motion
\begin{equation}
m_j \frac{d^2 {\bf r}_j}{d t^2} = \sum_{l \neq j}^4 \frac{G e_l e_j}{r_{lj}^3} ({\bf r}_l - {\bf r}_j)\, , \quad \mbox{(for $j$ = 1, 2, 3, 4)} \label{1}
\end{equation}
where ${\bf r}_j$ denotes the position vector of particle $j$ in 3D-space, $m_j$ is its positive mass, $G$ is the positive constant of universal gravitation, $r_{lj} = |{\bf r}_l - {\bf r}_j|$ is the distance between particles $j$ and $l$, and $e_j$ is the charge of particle $j$ such that $m_j = |e_j|$, with two possible choices of sign for the charge.

In the mathematical literature we have found some papers in the context of determining Four-Body central configurations that consider negative masses \cite{ro}, \cite{ce}, \cite{ji}, with a different equation of motion, a different definition of central configuration and without distinguishing between masses and charges. The difference with respect to the differential equation of motion used in those papers is stressed in an Appendix at the end of the paper, where, assuming the validity of Newton's second and third laws, the equation of motion used by those Authors is shown to produce, for a system of two masses of equal magnitude but different inertial and gravitational sign,  a rigid body which is self-accelerated under no external force, violating Newton's first law expressing that a body subject to no external forces moves with constant velocity.

In the following we quote some fundamental equations of classical mechanics that are essential in Physics \cite{tm}, \cite{js}, yet different for the approach in the mathematical cited papers above.

Equation of motion (1) expresses Newton's second law equating the positive inertial mass times the acceleration to Newton's gravitational force. This force obeys the action-reaction law  or Newton's third law: the force vector that particle $j$ exerts on particle $l$ is of equal magnitude and opposite sign to the force that particle $l$ exerts on particle $j$. As a consequence, the sum over all $j$ of the various equations of motion is the null vector
\begin{equation}
\sum_{j = 1}^4 m_j \frac{d^2 {\bf r}_j}{d t^2} = {\bf 0}\, . \label{2}
\end{equation}

The four bodies' masses $m_1$, $m_2$, $m_3$ and $m_4$ are positive and generally different in value, but some may be equal.

The total mass is
\begin{equation}
m = \sum_{j=1}^4 m_j\, , \label{3}
\end{equation}
The center of mass position is defined as
\begin{equation}
{\bf c} = \frac{1}{m} \sum_{j = 1}^4 m_j {\bf r}_j\, . \label{4}
\end{equation}
Equation (\ref{2}) implies
\begin{equation}
\frac{d^2 {\bf c}}{d t^2} = {\bf 0}\, . \label{5}
\end{equation}
which asserts that the center of mass moves with constant velocity.

With no loss of generality we assume in this paper that
\begin{equation}
\frac{d {\bf c}}{d t} = {\bf 0}\, , \quad {\bf c} = {\bf 0}\, ,\quad \sum_{j = 1}^4 m_j {\bf r}_j = {\bf 0}\, . \label{6}
\end{equation}
The center of mass is thus at the origin of the system of coordinates ${\bf r}_j$.

From equation (\ref{1}) the conservation of total energy $E$ follows, namely
\begin{equation}
\frac{1}{2} \sum_{j = 1}^4 m_j \frac{d {\bf r}_j}{d t} \cdot \frac{d {\bf r}_j}{d t} - \sum_{l < j}^4 \frac{G e_j e_l}{r_{lj}} = E\, . \label{7}
\end{equation}
In this expression the first term on the left hand side is the kinetic energy, involving the positive inertial masses, while the second term is the potential energy, which depends on the gravitational charges.

Equation (\ref{1}) also implies conservation of angular momentum
\begin{equation}
\frac{d}{d t} \sum_{j = 1}^4 m_j {\bf r}_j \times \frac{d {\bf r}_j}{d t} = {\bf 0}\, , \label{8}
\end{equation}
which again contains the inertial masses.

\section{Central configurations of 4 masses}
We define a central configuration as one in which the particle positions, defined up to rotation around the center of mass and uniform dilatations, obey the equation
\begin{equation}
B m_j {\bf r}_j = \sum_{l \neq j}^4 \frac{G e_l e_j}{r_{lj}^3} ({\bf r}_l - {\bf r}_j)\, , \quad \mbox{(for $j$ = 1, 2, 3, 4)}\, , \label{9}
\end{equation}
where $B$ is the same quantity for all the particles. In the right side we have the gravitational force acting on particle $j$ due to the other charges. Using the fact that the sum of forces is zero, we recover the third element of hypothesis (\ref{6}): the origin of coordinates ${\bf r}_j$ is at the center of mass.

The force of the right hand side of (\ref{9}) is the gradient of the potential energy
\begin{equation}
\sum_{l \neq j}^4 \frac{G e_l e_j}{r_{lj}^3} ({\bf r}_l - {\bf r}_j) = \frac{\partial}{\partial {\bf r}_j} \sum_{l < k}^4 \frac{G e_k e_l}{r_{lk}}\, . \label{10}
\end{equation}
The left hand side of (9) is related with the gradient of
the total moment of inertia, $I$,
\begin{equation}
m_j {\bf r}_j = \frac{\partial}{\partial {\bf r}_j}  \frac{1}{2} \sum_{k = 1}^4 m_k {\bf r}_k \cdot {\bf r}_k\, , \label{11}
\end{equation}

which may be written in terms of the relative distances as
\begin{equation}
I = \frac{1}{2} \sum_{k = 1}^4 m_k {\bf r}_k \cdot {\bf r}_k = \frac{1}{2 m} \sum_{k \neq l}^4 m_k m_l r_{kl}^2\, . \label{12}
\end{equation}
It follows that the condition for central configurations of four particles in three dimensions may be expressed in terms of derivatives with respect to the relative distances $r_{ij}$ in the form
\begin{equation}
\frac{e_l e_j}{r_{lj}^3} = \sigma m_l m_j \, , \label{13}
\end{equation}
where $\sigma$ is a quantity containing $B$ and $G$. The left hand side of this equation is the derivative of the potential energy with respect to the square of the distance $r_{lj}^2$, while the right hand side is proportional to the derivative of the total moment of inertia with respect to the same variable $r_{lj}^2$. But this equation can not possibly be satisfied for all six combinations of indexes because the left hand side takes different signs and the right hand side has the sign of $\sigma$ because the m's are positive. One concludes that there is no central configuration in 3-space for particles with different charges. Thus, the Lehman-Filh\'es equilateral tetrahedron  solution \cite{lf} valid for positive charges is not a central configuration for the case of positive and negative charges.

Consider now the question of possible planar configurations. The modified Dziobek equations for planar central configurations of positive and negative charges are
\begin{equation}
\frac{e_l e_j}{r_{lj}^3} = \sigma m_l m_j + \lambda S_l S_j\, , \label{14}
\end{equation}
where $\lambda$ is a new parameter and $S_k$ are the directed areas of the triangles having at their vertexes  the three particles different from $k$.  With respect to equation (\ref{13}), the additional term takes into account the zero volume restriction necessary for planar configurations: $S_l S_k$ is a constant times the derivative with respect to $r_{lk}^2$ of the square of the volume of the tetrahedron formed by the four particles, which equals the so-called Cayley-Menger determinant. A similar equation was obtained by Dziobek \cite{dz}, but although the two terms on the right hand side of this equation are the same as in Dziobek's paper, in the left hand side the charges are replaced by masses.

Note the invariance of the modified Dziobek's equations with respect to a sign change of all the charges, as well as their invariance with respect to a sign change of the directed areas
$$
e_j \quad \longrightarrow \quad - e_j \,, \quad S_j \quad \longrightarrow \quad - S_j \, .
$$

In order to prove that these equations are equivalent to the equations defining the central configurations, we need the planar conditions \cite{dz}, \cite{pl}
\begin{equation}
\sum_{j = 1}^4 S_j = 0 \, ,\quad \sum_{j = 1}^4 S_j {\bf r}_j = {\bf 0}\, . \label{15}
\end{equation}
Substituting the left hand side of (14) by the right hand side in (9), one obtains
\begin{equation}
B m_j {\bf r}_j = G \sum_{l \neq j}^4 ({\bf r}_l - {\bf r}_j) [\sigma m_l m_j + \lambda S_l S_j] = - G \sigma m m_j {\bf r}_j\, , \label{16}
\end{equation}
where we used the conditions (\ref{6}) that the center of mass is at the origin, and the planar configuration properties (\ref{15}). Thus from (\ref{16}) we obtain that
\begin{equation}
B = - G \sigma m\, . \label{17}
\end{equation}

In addition, one has the property
\begin{equation}
\sum_{l = 1}^4 \sum_{j = 1}^4 S_l S_k r_{lj}^2 = \sum_{l = 1}^4 \sum_{j = 1}^4 S_l S_j ({\bf r}_l^2 -2 {\bf r}_l \cdot {\bf r}_j + {\bf r}_j^2) = 0\, ,  \label{18}
\end{equation}
where we use the planar configuration conditions (\ref{15}). Note that the terms with $j =l$ in these summations are zero; therefore if we multiply both sides of equation (\ref{14}) by $r_{lj}^2$ and we sum over all $l$ and $j$ with $l \neq j$ one obtains
\begin{equation}
\sum_{l, j , l\neq j} \frac{e_l e_j}{r_{lj}} = \sigma \, \sum_{l, j , l\neq j} m_l m_j r_{lj}^2\, . \label{19}
\end{equation}
In this case $\sigma$ has not a definite sign because of the presence of  charges with both signs.

Write equation (\ref{14}) divided by the product of masses $m_l m_j$ as
\begin{equation}
\frac{e_l e_j}{m_l m_j} \frac{1}{r_{lj}^3} = \sigma + \lambda A_l A_j \, , \label{falta}
\end{equation}
where $A_j = S_j/ m_j$ denotes, as in \cite{pl}, weighted area. Noting that the square of $\frac{e_l e_j}{m_l m_j}$  equals positive one, this equation may also be written as
\begin{equation}
\left(\frac{e_l e_j}{m_l m_j} \frac{1}{r_{lj}}\right)^3 = \sigma + \lambda A_l A_j \, , \label{20}
\end{equation}
 A similar equation is the basic tool to compute central configurations in reference \cite{pl} from the given weighted areas $A_j$. The difference is the replacement of distance $r_{lj}$ by a sort of charged distance
 \begin{equation}
 r_{lj} \quad \longrightarrow \quad \frac{m_l m_j}{e_l e_j} r_{lj}\, . \label{21}
 \end{equation}
The algorithm presented in \cite{pl} is also useful for the present case if allowance is made for these positively or negatively charged distances in the numerical computation. In that paper the algorithm to compute planar central configurations starting from the weighted areas was applied to several examples with positive charges. Since the areas of the triangles having its vertices at the positions of the particles were computed by Heron's formula in terms of the square of this distances, they are not modified by the extra factor $(m_l m_j)/(e_l e_j)$. At the end of the present paper the modified algorithm is used to obtain some numerical examples of central configurations combining it with the new algorithm recently proposed to compute central configurations from given masses \cite{np} based on a new system of coordinates \cite{pi} which will be described in the next section. We refer the reader to those references for a more detailed account.

Substracting term by term equation (\ref{falta}) with subscripts $l, j$ and $l, k$, and with subscripts $n, j$ and $n, k$, with all subscripts different, yields
\begin{equation}
\frac{e_l e_j}{m_l m_j} \frac{1}{r_{lj}^3} - \frac{e_l e_k}{m_l m_k} \frac{1}{r_{lk}^3} = \lambda A_l (A_j - A_k)\, ,
\end{equation}
\begin{equation}
\frac{e_n e_j}{m_n m_j} \frac{1}{r_{nj}^3} - \frac{e_n e_k}{m_n m_k} \frac{1}{r_{nk}^3} = \lambda A_n (A_j - A_k)\, .
\end{equation}
Elimination of $\lambda$ between these two equations gives the fundamental relation
\begin{equation}
S_n e_l \left( \frac{e_j m_k}{r_{lj}^3} - \frac{e_k m_j}{r_{lk}^3} \right) = S_l e_n \left( \frac{e_j m_k}{r_{nj}^3} - \frac{e_k m_j}{r_{nk}^3} \right)\, . \label{fund}
\end{equation}
The same equation was obtained directly from two of the defining equations for planar central configurations (\ref{9}). A similar calculation may be found in reference \cite{pl}. From the above equation we prove with no loss of generality that the sign of the charge of any particle may be made to coincide with the sign of the corresponding area.

Now, assume that the charge of particle 1 is of opposite sign to the charge of the other three particles. Using equation (\ref{fund}) with $j=1$, we note that the quantity in the two parenthesis has the same sign and therefore
\begin{equation}
\frac{S_2}{S_3} > 0 \, , \quad \frac{S_3}{S_4} > 0 \, , \quad \frac{S_4}{S_2} > 0 \, .
\end{equation}
With no loss of generality we may choose
\begin{equation}
S_2 > 0 \, , \quad S_ 3 > 0 \, , \quad S_4 > 0 \, , \quad S_1 < 0 \, .
\end{equation}
Therefore one charge opposite in sign to the other three leads to a concave configuration with the different sign charge in the convex hull of the other three.

Consider now the case in which particles 1 and 2 have charges of one sign and particles 3 and 4 of the opposite sign.  Again using equation (\ref{fund}) with charge $j$  of one sign and charge $k$ of the opposite sign, the quantity in the parenthesis has the same sign, so that
\begin{equation}
\frac{S_1}{S_3} < 0 \, , \quad \frac{S_1}{S_4} < 0 \, , \quad \frac{S_2}{S_3} < 0 \, , \quad \frac{S_2}{S_4} < 0 \, .
\end{equation}
With no loss of generality we take
\begin{equation}
S_1 > 0 \, , \quad S_ 2 > 0 \, , \quad S_3 < 0 \, , \quad S_4 < 0 \, .
\end{equation}
Therefore two pairs of particles of different sign lead to a convex configuration with charges of the same sign located at the ends of the two diagonals.

As a consequence, the quotient $S_k/e_k$ may be considered to be always positive.

It is convenient to divide both members of equation (\ref{14}) by the product of charges $e_l e_j$ to yield
\begin{equation}
\frac{1}{r_{lj}^3} = \sigma \frac{m_l m_j}{e_l e_j} + \lambda \frac{S_l S_j}{e_l e_j}\, .
\end{equation}
Choosing two of these equations, with subscripts say  $l, j$ and $l, k$, such that the term containing $\sigma$ in each is of opposite sign, adding member by member we obtain
\begin{equation}
\left( \frac{1}{r_{lj}^3} + \frac{1}{r_{lk}^3} \right) = \lambda \left( \frac{S_l S_j}{e_l e_j} + \frac{S_l S_k}{e_l e_k}\right)\, .
\end{equation}
Since the quantities in the parentheses are both positive, we have proved that
\begin{equation}
\lambda > 0 \, .
\end{equation}

\section{New coordinates in the Four-Body Problem}

This section reviews the main ideas and results of the new four-body coordinates  of \cite{pi}, slightly expanded at a few spots but condensed to the minimum necessary.

We transform from the inertial referential system to the frame of principal axes of inertia by means of a three-dimensional rotation $\bf G$ parameterized by three independent coordinates. In addition to this rotation, three more coordinates are introduced, as scale factors $R_1$, $R_2$, $R_3$, which are three directed distances closely related to the three principal moments of inertia through
\begin{equation}
I_1 = \mu (R_2^2 + R_3^2)\, , \quad I_2 = \mu (R_3^2 + R_1^2) \, , \quad \mbox{and} \quad I_3 = \mu (R_1^2 + R_2^2) \, , \label{22}
\end{equation}
where $\mu$ is the mass
\begin{equation}
\mu = \sqrt[3]{\frac{m_1 \, m_2\, m_3\, m_4}{m_1 + m_2 + m_3 + m_4}} \, . \label{23}
\end{equation}

The first rotation matrix rotates to the principal axes of inertia and after the scale change the resulting four-body configuration has a moment of inertia tensor with the three principal moments of inertia equal. The second rotation $\bf G'$ does not change this property.

The cartesian coordinates of the four particles, with the origin at the center of gravity, written in terms of the new coordinates are
\begin{equation}
\left(\begin{array}{cccc}
x_1 & x_2 & x_3 & x_4 \\
y_1 & y_2 & y_3 & y_4 \\
z_1 & z_2 & z_3 & z_4
\end{array} \right) = {\bf G} \left(\begin{array}{ccc}
R_1 & 0 & 0 \\
0 & R_2 & 0 \\
0 & 0 & R_3
\end{array} \right) {\bf G'}^{\rm T}
\left(\begin{array}{cccc}
a_1 & a_2 & a_3 & a_4 \\
b_1 & b_2 & b_3 & b_4 \\
c_1 & c_2 & c_3 & c_4
\end{array} \right)\, , \label{24}
\end{equation}
where $\bf G$ and $\bf G'$ are two rotation matrices, each a function of three independent coordinates such as the Euler angles, and where the column elements of the constant matrix
\begin{equation}
{\bf E} = \left(\begin{array}{cccc}
a_1 & a_2 & a_3 & a_4 \\
b_1 & b_2 & b_3 & b_4 \\
c_1 & c_2 & c_3 & c_4
\end{array} \right)\, , \label{25}
\end{equation}
are the coordinates of the four vertexes of a rigid orthocentric tetrahedron \cite{co}, having its center of mass at the origin of coordinates, namely:
\begin{equation}
\begin{array}{c}
a_1 m_1 + a_2 m_2 + a_3 m_3 + a_4 m_4 = 0 \, , \\
b_1 m_1 + b_2 m_2 + b_3 m_3 + b_4 m_4 = 0 \, , \\
c_1 m_1 + c_2 m_2 + c_3 m_3 + c_4 m_4 = 0 \ ,
\end{array}\, . \label{base}
\end{equation}

We introduce the mass matrix
\begin{equation}
{\bf M} = \left( \begin{array}{cccc}
m_1 & 0 & 0 & 0 \\
0 & m_2 & 0 & 0 \\
0 & 0 & m_3 & 0 \\
0 & 0 & 0 & m_4
\end{array} \right) \, . \label{26}
\end{equation}

An equivalent condition for having three equal moments of inertia for the rigid tetrahedron is expressed as
\begin{equation}
{\bf E\, M\, E}^{\rm T} = \mu \left( \begin{array}{ccc}
1 & 0 & 0 \\
0 & 1 & 0 \\
0 & 0 & 1
\end{array} \right) \label{ort}
\end{equation}

The coordinate system for measuring the $\bf G'$ rotation can be chosen in various ways, from which we prefer to use the same coordinates as in reference \cite{np}, namely, particle with mass $m_1$ along coordinate axis 3, the other three in a  plane parallel to the coordinate plane containing axes 1 and 2 but that does not include the particle of mass $m_1$; the particle with mass $m_2$ on an orthogonal coordinate plane that contains the first particle and the center of mass, and the other two particles on a line that is parallel to coordinate axis 1 and perpendicular to the coordinate plane containing the first two particles. Particle 1 thus has coordinates
\begin{equation}
(a_1, b_1, c_1) = \left(0, 0, \sqrt{\frac{\mu (m - m_1)}{m_1 m}}\right)\, . \label{29}
\end{equation}
Particle 2 has coordinates
\begin{equation}
(a_2, b_2, c_2) = \left(0, \sqrt{\frac{\mu (m_3+m_4)}{m_2 (m - m_1)}}, - \sqrt{\frac{\mu m_1}{(m - m_1) m}}\right)\, .
\end{equation}
Particle 3 has coordinates
$$
(a_3, b_3, c_3) =
$$
\begin{equation}
\left(\sqrt{\frac{\mu m_4}{m_3 (m_3+m_4)}}, - \sqrt{\frac{\mu m_2}{(m_3+m_4) (m - m_1)}}, - \sqrt{\frac{\mu m_1}{(m - m_1) m}}\right)\, .
\end{equation}
Particle 4 has coordinates
$$
(a_4, b_4, c_4) =
$$
\begin{equation}
\left(- \sqrt{\frac{\mu m_3}{m_4 (m_3+m_4)}}, - \sqrt{\frac{\mu m_2}{(m_3+m_4) (m - m_1)}}, - \sqrt{\frac{\mu m_1}{(m - m_1) m}}\right)\, . \label{32}
\end{equation}

Note that this election of coordinate axes implies $a_1=a_2=0$, $b_3=b_4$ and $c_2=c_3=c_4$.

This rigid tetrahedron is the generalization of the rigid triangle of the Three-Body problem with the center of mass at the orthocenter discussed previously in \cite{pb}. The same triangle was used with different purposes by C. Simo \cite{si}.

Our coordinates are now adapted to planar configurations, with the four particles in a constant plane which may be chosen as the $z=0$ plane. Because of this fact, the first rotation is just by one angle in the plane of motion and the scale factor associated with the third coordinate is zero, namely
$$
\left(\begin{array}{cccc}
x_1 & x_2 & x_3 & x_4 \\
y_1 & y_2 & y_3 & y_4 \\
0 & 0 & 0 & 0
\end{array} \right) =
$$
\begin{equation}
\left(\begin{array}{ccc}
\cos \psi & -\sin \psi & 0\\
\sin \psi & \cos \psi & 0 \\
0 & 0 & 1
\end{array} \right)
 \left(\begin{array}{ccc}
R_1 & 0 & 0 \\
0 & R_2 & 0 \\
0 & 0 & 0
\end{array} \right) {\bf G'}^{\rm T}
\left(\begin{array}{cccc}
a_1 & a_2 & a_3 & a_4 \\
b_1 & b_2 & b_3 & b_4 \\
c_1 & c_2 & c_3 & c_4
\end{array} \right)\, .
\end{equation}
This equation simplifies to
$$
\left(\begin{array}{cccc}
x_1 & x_2 & x_3 & x_4 \\
y_1 & y_2 & y_3 & y_4
\end{array} \right) =
$$
\begin{equation}
\left(\begin{array}{cc}
\cos \psi & -\sin \psi \\
\sin \psi & \cos \psi
\end{array} \right) \left(\begin{array}{ccc}
R_1 & 0 & 0 \\
0 & R_2 & 0
\end{array} \right) {\bf G'}^{\rm T}
\left(\begin{array}{cccc}
a_1 & a_2 & a_3 & a_4 \\
b_1 & b_2 & b_3 & b_4 \\
c_1 & c_2 & c_3 & c_4
\end{array} \right)\, , \label{flat}
\end{equation}
in terms of six degrees of freedom.

The corresponding expression for the four directed areas in terms of these coordinates is
\begin{equation}
\left( \begin{array}{c}
S_1 \\
S_2 \\
S_3 \\
S_4
\end{array} \right) =  C {\bf M E}^{\rm T} {\bf G'} \left( \begin{array}{c}
0 \\
0 \\
1
\end{array} \right) \, , \label{areas}
\end{equation}
where $C$ is a constant with units of area divided by mass. Note that Equations (\ref{15}) are satisfied from this expression of the directed areas since substitution of equations (\ref{flat}) and (\ref{areas}) in equations (\ref{15}), and application of equation (\ref{ort}), yields an identity, independent of coordinates.

Out of the three angles necessary to describe arbitrary rotations in space, for $\bf G'$ in (\ref{areas}) just two rotation angles are needed, for which we choose those required to express the unit vector in spherical coordinates
\begin{equation}
{\bf G'} \left( \begin{array}{c}
0 \\
0 \\
1
\end{array} \right) = \left( \begin{array}{c}
\sin \theta \cos \phi \\
\sin \theta \sin \phi \\
\cos \theta
\end{array} \right)\, ,
\end{equation}
where $\theta$ and $\phi$ are the spherical coordinates determining this vector.
Given the four masses, the four directed areas are functions, up to a multiplicative constant $C$ depending on the choice of physical units, of this unit vector direction only.

The non-collinear planar central configurations are characterized in our coordinates by constant values of the $\bf G'$ matrix and of the constant value of the ratio $R_1/R_2$, which are not arbitrary, but are determined by three independent quantities as discussed in the following.

From (\ref{areas}) follows that in a planar solution the weighted directed areas are expressed as
\begin{equation}
\left( \begin{array}{c}
A_1 \\
A_2 \\
A_3 \\
A_4
\end{array} \right) =  C {\bf E}^{\rm T} {\bf G'} \left( \begin{array}{c}
0 \\
0 \\
1
\end{array} \right) = C {\bf E}^{\rm T} \left( \begin{array}{c}
\sin \theta \cos \phi \\
\sin \theta \sin \phi \\
\cos \theta
\end{array} \right)\, . \label{sec}
\end{equation}
Therefore, the weighted directed areas are up to a normalization factor equal to the third rotated coordinate of the rigid tetrahedron. In terms of the vectors (\ref{29}-\ref{32}) and the angles $\theta$ and $\phi$ this equation is expressed simply as
\begin{equation}
Aj = C (a_j \sin \theta \cos \phi + b_j \sin \theta \sin \phi + c_j \cos \theta) \, .
\end{equation}
Choosing $C = \sqrt{(m-m_1)/\mu}$ we have explicitly
$$
A_1 = \frac{m - m1}{\sqrt{m_1 m}} \cos \theta \, ,
$$
$$
A_2 = - \sqrt{\frac{m_1}{m}} \cos \theta + \sqrt{\frac{m_3 + m_4}{m_2}} \sin \theta \sin \phi \, ,
$$
$$
A_3 = - \sqrt{\frac{m_1}{m}} \cos \theta - \sqrt{\frac{m_2}{m_3 + m_4}} \sin \theta \sin \phi + \sqrt{\frac{m_4 (m - m_1)}{m_3 (m_3 + m_4)}} \sin \theta \cos \phi
$$
\begin{equation}
A_4 = - \sqrt{\frac{m_1}{m}} \cos \theta - \sqrt{\frac{m_2}{m_3 + m_4}} \sin \theta \sin \phi - \sqrt{\frac{m_3 (m - m_1)}{m_4 (m_3 + m_4)}} \sin \theta \cos \phi\, .\label{sfero}
\end{equation}
Note that a sign change of the unit vector in (47) produces a simultaneous sign change in these four $A_j$'s which does not to give a different central configuration. Therefore, it suffices to consider only the hemisphere $0 \leq \theta \leq \pi/2$.

\section{Computing central configurations for positive and negative charges}
In the case of positive charges, since the lengths and masses are defined up to arbitrary units, with no loss of generality we assumed  in \cite{pl} that the parameter $\sigma$ equals plus one. However, central configurations with charges of different sign obtained numerically always yield a negative $\sigma$, so that for those cases we used $\sigma=-1$  as follows
\begin{equation}
\frac{e_j\, e_k}{m_j\, m_k} \, r_{jk}^{-3} = - 1 + \lambda A_j A_k\, \quad (j \neq k).
\end{equation}

We recall that in the paper by Pi\~na and Lonngi \cite{pl} the assumption that the directed weighted areas are known as four given constants was made. The equation which corresponds to (51) then gives the distances as functions of the unknown parameter $\lambda$. Through them, the areas of the four triangles become functions of $\lambda$, that should satisfy restrictions (\ref{15}) for a planar solution. These restrictions allow in many cases to determine the value of $\lambda$ and hence the values of the six distances and the four masses. This is an implicit way to deduce planar central configurations with four masses.

In contrast, in this paper, as well as in reference \cite{np}, we assume that the four masses are known from the beginning. The four weighted areas are then determined by expressions (\ref{sfero}) in terms of the two tuning variables $\theta$ and $\phi$. Particular values of these two angles determine the four constants $A_j$, up to a multiplicative factor, which in turn produce a central configuration with computed distances and masses. The computed masses are in general not equal (or proportional) to the starting values used to build the orthocentric tetrahedron. The two angles are then tuned until a numerical match is produced between the given and the computed masses. The distances between particles, computed for this central configuration, correspond to the given masses.

\begin{figure}
\centering
\scalebox{0.6}{\includegraphics{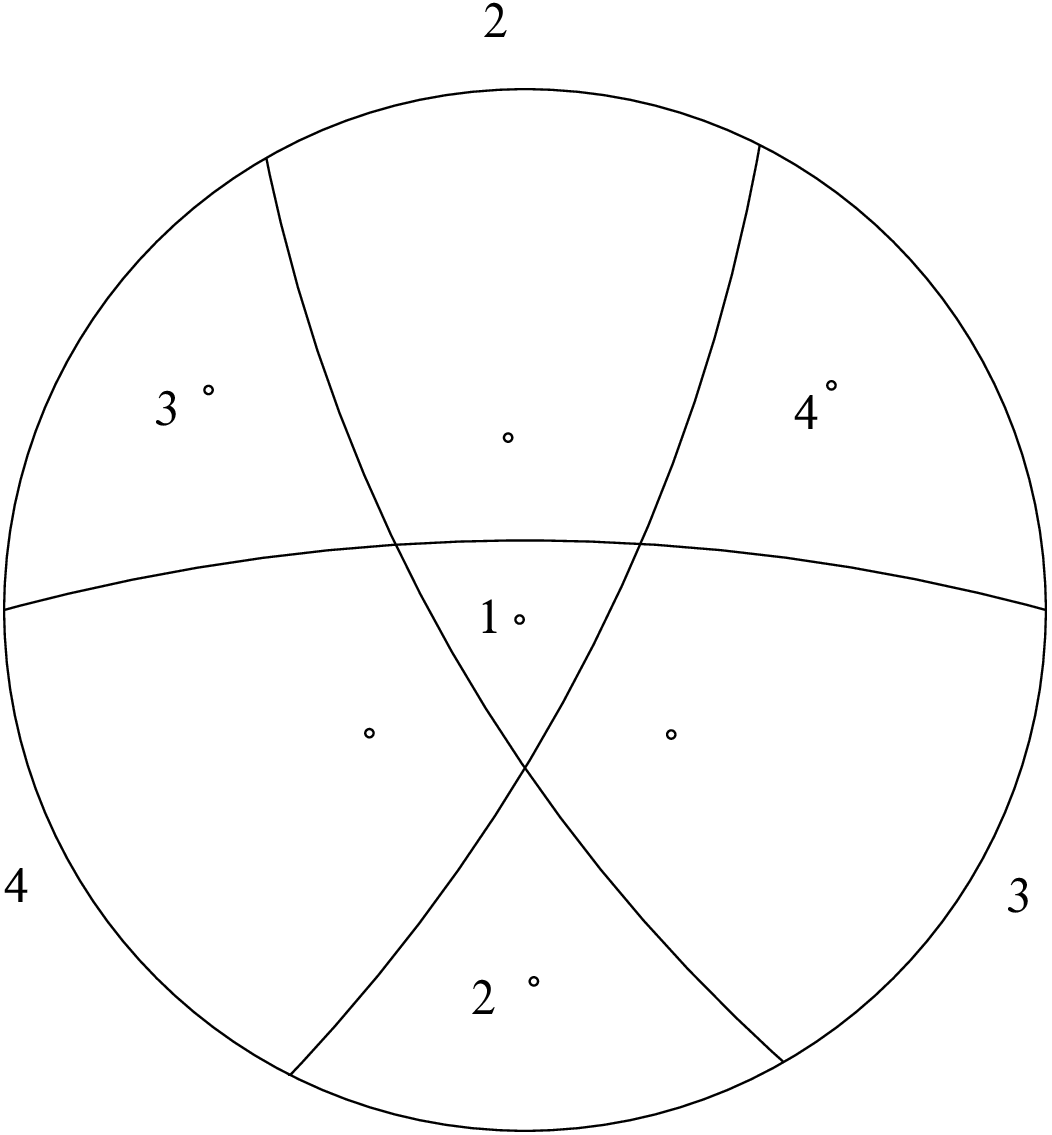}}

\caption{Stereographic projection of the hemisphere of the two angles motion of the orthocentric tetrahedron. The great circles represent the positions where three particles are collinear. The four spherical triangles are concave open sets labeled by the particle at the interior of the triangle. The spherical quadrilateral open sets correspond to convex configurations with the same order that the neighboring triangles. The isolated points are at the angles where a charged central configuration has been computed. The values of the masses are $m_1=10, m_2=13, m_3=15, m_4=17$. The sign of the charges is opposite for the position inside a concave spherical triangle set and the same for the two particles along each diagonal when the position is inside a rectangular convex region.}
\end{figure}

For the arbitrarily chosen values of the masses  $m_1 = 10, m_2 = 13, m_3 = 15, m_4 = 17$  seven central configurations are found, as follows:

\noindent A concave one with $e_1 = - m_1$ at the interior, for $\lambda = 47.2854408881827$
$$
\begin{array}{ll}
\theta = 0.042093321601107 & \phi = 4.19675592202334 \\
r_{31} = 0.292697411518074 & r_{42} = 0.506907815953173 \\
r_{41} = 0.304528873642513 & r_{23} = 0.484280324121928 \\
r_{12} = 0.281506710172786 & r_{43} = 0.53050452673262
\end{array}
$$
A concave one with $e_2 = - m_2$ at the interior, for $\lambda = 12.0640816044133$
$$
\begin{array}{ll}
\theta = 1.23894072035608 & \phi = 4.7362647130086 \\
r_{23} = 0.454852343412694 & r_{41} = 0.756907150900389 \\
r_{42} = 0.470894058612881 & r_{31} = 0.717209067230588 \\
r_{12} = 0.42279600672246 & r_{43} = 0.854639094313928
\end{array}
$$
A concave one with $e_3 = - m_3$ at the interior, for $\lambda = 7.82519752909191$
$$
\begin{array}{ll}
\theta = 1.22013530709446 & \phi = 2.62428566877055 \\
r_{23} = 0.506175567822397 & r_{41} = 0.892326241528415 \\
r_{43} = 0.539174524504944 & r_{12} = 0.793305326469069 \\
r_{31} = 0.488136474707177 & r_{42} = 0.961093408384294
\end{array}
$$
A concave one with $e_4 = - m_4$ at the interior, for $\lambda = 5.66837976372513$
$$
\begin{array}{ll}
\theta = 1.20553148849283 & \phi = 0.545338011287725 \\
r_{41} = 0.542962598397597 & r_{23} = 1.04170674195263 \\
r_{43} = 0.577658606056266 & r_{12} = 0.902271887672873 \\
r_{42} = 0.561346152785465 & r_{31} = 0.961453082272945
\end{array}
$$
A convex one with $r_{41}, r_{23}$ at the diagonals, different charges on each diagonal, for $\lambda = 3.01670996202888$
$$
\begin{array}{ll}
\theta = 0.714310336851875 & \phi = 3.78435778489043 \\
r_{41} = 0.930768749839807 & r_{23} = 0.97553708802784 \\
r_{12} = 0.580828657475032 & r_{43} = 0.776730715251531 \\
r_{31} = 0.619329983908135 & r_{42} = 0.741807873791024
\end{array}
$$
A convex one with $r_{42}, r_{31}$ at the diagonals, different charges on each diagonal, for $\lambda = 3.206938002701911$
$$
\begin{array}{ll}
\theta = 0.693611387046597 & \phi = 5.60199455350564 \\
r_{42} = 1.02758305520169 & r_{31} = 0.84981686247219 \\
r_{12} = 0.565145699097753 & r_{43} = 0.779304949192135 \\
r_{41} = 0.632347185116331 & r_{23} = 0.717265794360609
\end{array}
$$
A convex one with $r_{43}, r_{12}$ at the diagonals, different charges on each diagonal, for $\lambda = 3.53428154586715$
$$
\begin{array}{ll}
\theta = 0.641643146536123 & \phi = 1.66856071624655 \\
r_{43} = 1.05828492646533 & r_{12} = 0.766407979780657 \\
r_{31} = 0.572866264461101 & r_{42} = 0.749501180861624 \\
r_{41} = 0.602822523027629 & r_{23} = 0.721181661002597
\end{array}
$$
We recall that, using the same values for the masses, seven different central configurations were found for positive charges \cite{np}, each with its own values for the angles $\theta$ and $\phi$. On the other hand, allowing for negative charges, we only have one central configuration for each possible election of one negative charge and of pairs of charges of the same sign, with angles different from those previously computed in \cite{np}. Thus, the possibility of charges of different sign reduces the number of different central configurations to just one for each election.

\section*{Appendix}
In this paper we assume an equation of motion for the particles different from that used in references \cite{ro}, \cite{ce}, \cite{ji}. We will show that their equation of motion applied to two particles of opposite charge, leads to a contradiction with  Newton's  first law of motion: under no external force, a body moves with a constant velocity vector.

To help distinguish clearly, we continue using the notation $e_j$ to denote the charge with two possible choices of sign which these Authors denote with $m_j$. Their equation of motion is
\begin{equation}
\frac{d^2 {\bf r}_1}{d t^2} = \frac{G e_2}{r_{12}^3} ({\bf r}_2 - {\bf r}_1)\, , \quad \frac{d^2 {\bf r}_2}{d t^2} = \frac{G e_1}{r_{12}^3} ({\bf r}_1 - {\bf r}_2)\, , \label{A1}
\end{equation}
which differ from equation (\ref{1}) due to substitution of the mass $m_j$ by the charge $e_j$, and that only two particles of opposite charge $e_1 = - e_2$ are considered.

The product of the first times $e_1$, of the second times $e_2$ and adding member by member produces instead of zero acceleration of the center of mass, zero acceleration of the relative position
\begin{equation}
e_1 \frac{d^2 ({\bf r}_2 - {\bf r}_1)}{d t^2} = {\bf 0}\, ,
\end{equation}
which is trivially integrated in terms of two constant vectors of integration $\bf A$ and $\bf B$
\begin{equation}
{\bf r}_2 - {\bf r}_1 = {\bf A} + {\bf  B} t\, .
\end{equation}

Assume the initial condition ${\bf B} = {\bf 0}$. Then the relative distance is constant
\begin{equation}
{\bf r}_2 - {\bf r}_1 = {\bf A}\, ,
\end{equation}
as in a rigid body. The two particles are however accelerated with the same constant acceleration
\begin{equation}
\frac{d^2 {\bf r}_1}{d t^2} = \frac{d^2 {\bf r}_2}{d t^2} = \frac{G e_2}{|{\bf A}|^3} {\bf A}\, ,
\end{equation}
which contradicts Newton's first law since no external force is acting on this composite body.


\begin{thebibliography}{aa}
\bibitem{tm} S. T. Thornton \& J. B. Marion \textit{Classical Dynamics of Particles and Systems} 5th edition (Thomson Brooks/Cole, Mexico, 2004).

\bibitem{js} J. V. Jose \& E. J. Saletan \textit{Classical Mechanics, A Contemporary Approach} (Cambridge University Press, Cambridge, 1998).

\bibitem{ro} G. E. Roberts, \textit{A continuum of relative equilibria in the five-body problem} Physica D \textbf{127} 141-145 (1999).

\bibitem{ce} M. Celli, \textit{The central configurations of four masses $x$, $-x$, $y$, $-y$} J. Differential Equations \textbf{235} 668-682 (2007).

\bibitem{ji} J. Shi, Z. Xie \textit{Classification of four-body central configurations with three equal masses} J. Mat. Anal Appl. \textbf{363} 512-524 (2010).

\bibitem{lf} R. Lehman-Filh\'es, \textit{Ueber swei F\"alle des Vielk\"orpersproblems} Astr. Nachr. \textbf{127} 137-144 (1891).

\bibitem{dz} O. Dziobek, \textit{\"Uber einer werkw\"urdingen Fall des Vielk\"orperproblems} Astr. Nachr. \textbf{152} 33-46 (1900).

\bibitem{pl} E. Pi\~na and P. Lonngi \textit{Central configurations for the planar Newtonian Four-Body Problem} Cel. Mech. \& Dyn. Astr. \textbf{108} 73-93 (2010).

\bibitem{np} E. Pi\~na, \textit{Algorithm for planar 4-Body Problem central configurations with given masses} ArXiv 1006.2430 (2010)

\bibitem{pi} E. Pi\~na, \textit{New coordinates for the Four-Body Problem} Rev. Mex. Fis. \textbf{56} 195-203 (2010).

\bibitem{co} N. A. Court \textit{Notes on the orthocentric tetrahedra}, The American Mathematical Monthly \textbf{41} (1934) 499-502

\bibitem{pb} E. Pi\~na and A. Bengochea \textit{Hyperbolic Geometry for the Binary Collision Angles of the Three-Body Problem in the Plane}, Qualitative Theor. of Dyn. Sys. Vol. 8, Pags. 399-417 (2009). Published online (09 february 2010).

\bibitem{si} C. Sim\'o El conjunto de bifurcaci\'on en el problema espacial de tres cuerpos. In: Acta I Asamblea Nacional de Astronom\'\i a y Astrof\'\i sica. Instituto de Astrof\'\i sica. (Univ. de la Laguna. Spain, 1975) pp. 211-217

\end{thebibliography}
\end{document}